\documentclass[english,aps,prl,twocolumn,superscriptaddress]{revtex4}
\usepackage{epsfig}
\usepackage{graphicx}
\usepackage{amsmath}
\usepackage{xcolor}
\usepackage[normalem]{ulem}
\usepackage{babel}
\usepackage{gensymb}

\begin{document}

\title{Peeling off neutron skins from neutron-rich nuclei:\\
Constraints on the symmetry energy from neutron-removal cross sections}

\author{T. Aumann}
\affiliation{Institut f\"ur Kernphysik, Technische Universit\"at Darmstadt, Schlossgartenstra\ss{}e 9, D-64289 Darmstadt, Germany}
\affiliation{GSI Helmholtzzentrum f\"ur Schwerionenforschung, Planckstr. 1, D-64291 Darmstadt, Germany}

\author{C.A. Bertulani}
\affiliation{Department of Physics and Astronomy, Texas A\&M University-Commerce, Commerce, TX 75429-3011, USA}
\affiliation{Institut f\"ur Kernphysik, Technische Universit\"at Darmstadt, Schlossgartenstra\ss{}e 9, D-64289 Darmstadt, Germany}

\author{F. Schindler}
\affiliation{Institut f\"ur Kernphysik, Technische Universit\"at Darmstadt, Schlossgartenstra\ss{}e 9, D-64289 Darmstadt, Germany}

\author{S. Typel}
\affiliation{Institut f\"ur Kernphysik, Technische Universit\"at Darmstadt, Schlossgartenstra\ss{}e 9, D-64289 Darmstadt, Germany}
\affiliation{GSI Helmholtzzentrum f\"ur Schwerionenforschung, Planckstr. 1, D-64291 Darmstadt, Germany}

\begin{abstract}
An experimentally constrained equation of state of neutron-rich matter is fundamental for the physics of nuclei and the astrophysics of neutron stars, mergers, core-collapse supernova explosions and the synthesis of heavy elements. To this end, we investigate the potential of constraining the density dependence of the symmetry energy close to saturation density through measurements of neutron-removal cross sections in high-energy nuclear collisions of a few hundreds MeV/nucleon to GeV/nucleon. We show that the sensitivity of the total neutron-removal cross section is high enough so that the required accuracy can be reached experimentally with the recent developments of new detection techniques. We quantify two crucial points to minimize the model dependence of the approach and to reach the required accuracy: the contribution to the cross section from inelastic scattering has to be measured separately in order to allow a direct comparison of experimental cross sections to theoretical cross sections based on density functional theory and eikonal theory. The accuracy of the reaction model should be investigated and quantified by the energy and target dependence of various nucleon-removal cross sections. Our calculations explore the dependence of neutron-removal cross sections on the neutron skin of medium-heavy neutron-rich nuclei, and we demonstrate that the slope parameter L of the symmetry energy could be constrained down to $\pm10$\,MeV by such a measurement with a 2\% accuracy of the measured and calculated cross sections.
\end{abstract}

\maketitle

The knowledge of the equation of state (EoS) of neutron-rich nuclear matter is fundamental for understanding the properties of neutron stars, such as their masses, radii, and gravitational-wave signatures, as well as the mechanisms of core-collapse supernovae and neutron-star mergers \cite{Gle97,Web05,Hae07,Lat12,LP04,BP12,Heb13}. Although the size of atomic nuclei differs by many orders of magnitude compared to neutron stars, their bulk properties, limits of stability, as well as the characteristics of collisions among heavy ions are governed by the same fundamental interaction. The atomic nucleus thus represents the natural and only environment to investigate nuclear matter in the laboratory, and constraints on the EoS can be obtained from measurements of bulk properties of neutron-rich nuclei.

The EoS of asymmetric nuclear matter is usually characterized by the symmetry energy $E_{sym}(\rho)$ with its value $J$ and slope $L=3\rho_0\delta E_{sym}(\rho)/\delta\rho |_{\rho_0}$ at saturation density $\rho_0$. In particular, the latter quantity is very poorly constrained experimentally. This is apparent when inspecting the various interactions in Hartree Fock (HF)  \cite{Dut12,Ben03,Sto07,Sel14} and relativistic mean-field (RMF) \cite{Bro00,TB01,Dut14,Ohn17}  models which have been adjusted to properties of nuclei, e.g., masses and charge radii \cite{Ber09}.  Although well calibrated forces describe ground-state properties of nuclei and their excitations satisfatorily, they exhibit a wide scatter in the $L$ parameter in a range of almost 0 to 150~MeV \cite{Roc11}. 

In recent years, two nuclear observables have been identified to potentially provide tight constraints on $L$ if accurately determined. These are the neutron-skin thickness $\Delta r_{np}$ of neutron-rich nuclei and the ground-state dipole polarizability $\alpha_D$. The connection between $\Delta r_{np}$ and properties of the neutron EoS has first been pointed out and quantified in Ref.\ \cite{Bro00}. A clear relation between the derivative of the neutron EoS close to saturation density and $\Delta r_{np}$ of $^{208}$Pb calculated with both relaticistic and non-relativistic models  was obtained \cite{Bro00,TB01}. This implies that a precise determination of the skin thickness would provide constraints on the density dependence of the neutron matter EoS or, equivalently, on the slope parameter $L$ of the symmetry energy. A similar relation is observed with the dipole polarizability as pointed out first in Ref.\ \cite{RNa10}. First measurements of this observable have been performed and the most precise value so far has been extracted for  $^{208}$Pb, where the measurements at RCNP were analyzed together with the world data set\,\cite{Tam11}. Including a result for the neutron-rich nucleus $^{68}$Ni\,\cite{Ros13}, the corresponding range of the slope $L$ lies between 20 and 66 MeV according to the analysis performed in Ref.\ \cite{Roc15}. One would need to measure the polarizability with better than 5\% uncertainty for neutron-rich nuclei, to reach the precision achieved for stable nuclei.

The experimental determination of $\Delta r_{np}$ is rather challenging, in particular for short-lived neutron-rich nuclei where this effect becomes most pronounced. The supposedly cleanest probe to obtain information on $\Delta r_{np}$ is electron scattering. The PREX experiment \cite{Abra12}, to be performed at JLAB, will measure the parity-violating asymmetry for $^{208}$Pb at a relatively low momentum transfer $q$, which is related to the neutron radius  \cite{Roc11}. The intended PREX precision for the final production run of $\pm 3$\% for the asymmetry translates into an uncertainty of $\Delta r_{np}$ of $\pm 0.06$ fm, and a constraint on $L$ to $\pm 40$ MeV \cite{Roc11}. 

In this Letter we propose to use total neutron-removal cross sections $\sigma_{\Delta N}$ in high-energy nuclear collisions (0.4 to 1\,GeV/nucleon) with secondary beams of neutron-rich nuclei with hydrogen and carbon targets as an alternative method. We will show that $\sigma_{\Delta N}$ is rather sensitive to the neutron-skin thickness and to the slope parameter $L$. The constraint will be derived similarly as for the measurement of the asymmetry at one momentum transfer $q$ in the case of PREX, namely by relating $\sigma_{\Delta N}$ calculated on the basis of proton and neutron point densities from density functional theory with the corresponding $L$ parameter of the respective functional. The scatter of theory points will provide an estimate of the model dependence of such an analysis. Following the same analysis as discussed in Ref.\ \cite{Roc11}, we have concluded that this method could potentially constrain $L$ to $\pm 10$ MeV if experiments could provide the related observable with the corresponding accuracy. The obvious advantage is the abundant number of events one can accumulate in facilities using hadronic collisions. This opens a new window of opportunity for future experiments in high-energy radioactive-beam facilities with the purpose to reveal the neutron skin of stable and unstable nuclear isotopes.

\begin{figure}[t]
\includegraphics[width=0.99\columnwidth]{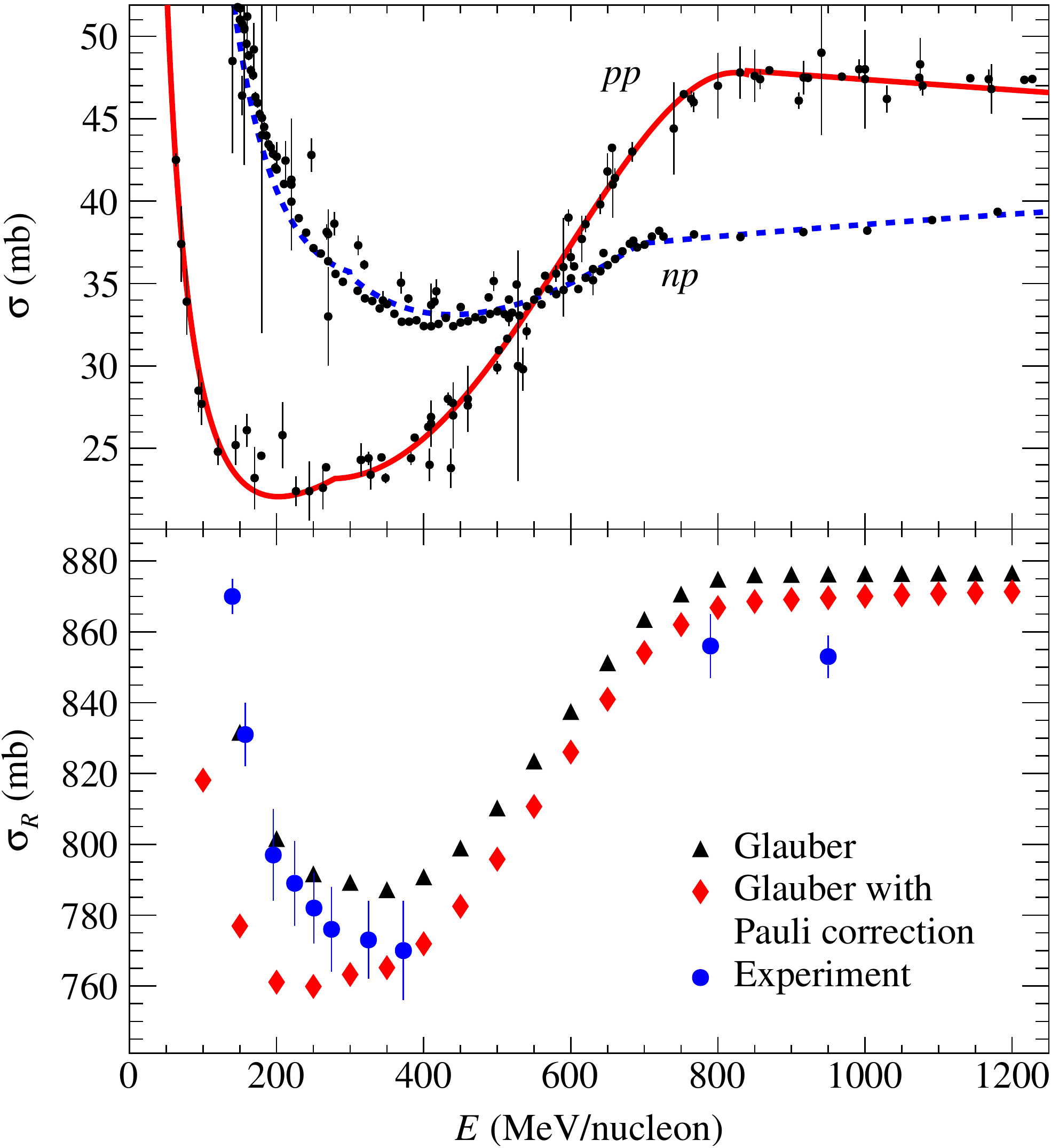}
\caption{(color online) Nucleon-nucleon (top) and total reaction cross sections for $^{12}$C on $^{12}$C (bottom) as a function of beam energy. The blue points display data from Refs.\ \cite{Tak09} ($100$ to $400$ MeV/nucleon), \cite{Tan90} (790 MeV/nucleon), and  \cite{Oza01} (950 MeV/nucleon). Black triangles display the result from a parameter-free eikonal calculation in the optical limit, while the red diamonds include the effect of Pauli blocking \cite{Schi17}.}
\label{fig1}
\end{figure}

Before discussing the sensitivity of $\sigma_{\Delta N}$ to $L$ and $\Delta r_{np}$ we briefly introduce our reaction model. In high-energy collisions, the Glauber multiple scattering method has been shown to be a reliable theoretical model to calculate the removal of nucleons \cite{BD04,Mill07}. The cross section for the production of a fragment $(Z,N)$ from a projectile $(Z_P,N_P)$  due to nucleon-nucleon collisions is given by 
\begin{eqnarray}
\sigma&=&
\left(
\begin{array}{c}Z_P \\ Z
\end{array}
\right)
\left(
\begin{array}{c}N_P \\ N
\end{array}
\right)
\int d^2 b \left[ 1-P_p(b)\right]^{Z_P-Z}P_p^Z(b) \nonumber \\
&\times& \left[ 1-P_n(b)\right]^{N_P-N}P_n^N(b), \label{sigma}
\end{eqnarray} 
where $b$ is the collision impact parameter and the binomial coefficients account for all possible combinations to select $Z$ protons out of the original $Z_P$ projectile protons, and similarly for the neutrons \cite{BD04,Mill07}. The probabilities for single nucleon survival are given by $P_p$ for protons and $P_n$ for neutrons, with the probability that a proton does not collide with the target given by  \cite{BD04,Mill07}
\begin{eqnarray}
P_p(b)&=&\int dzd^2s \rho_p^P({\bf s},z) \exp\left[ -\sigma_{pp} Z_T\int d^2s \rho_p^T({\bf b-s},z) \right. \nonumber \\
&-&\left. \sigma_{pn} N_T\int d^2s \rho_n^T({\bf b-s},z) \right],  \label{ppb}
\end{eqnarray} 
where $\sigma_{pp}$ and $\sigma_{np}$ are the proton-proton (Coulomb removed) and proton-neutron total cross sections, obtained from a fit of experimental data in the energy range of $10$ to $5000$\,MeV as in Eqs.\ (1) and (2) of Ref.\,\cite{BC10} (see Fig.\,\ref{fig1}).
The projectile (target) proton (neutron) densities are given by $\rho_{n(p)}^{P(T)}$ for the proton and neutron point densities in the projectile and in the target, respectively. They are normalized so that $\int d^3r\rho^{P(T)}_{n(p)} (r) = 1$. The expression for $P_n$ is similar to Eq.\,\eqref{ppb} with the replacement $n \leftrightarrow p$.  As the nucleon-nucleon cross sections are taken from experiment, the only input parameters in this model are the nuclear proton and neutron densities which can be directly taken from density functional theory and tested in comparison with the experimental cross sections. We will concentrate on the total neutron-removal, $\sigma_{\Delta N}$, charge-changing,  $\sigma_{\Delta Z}$, and reaction cross sections, $\sigma_{R}=\sigma_{\Delta N}+\sigma_{\Delta Z}$. These are obtained from a sum of all corresponding fragments using Eqs. \,\eqref{sigma} and \eqref{ppb}. 

We choose the neutron-rich part of the tin isotopic chain for our investigations and concentrate on the reactions Sn+$^{12}$C first. For $^{12}$C we adopt the density derived in a model-independent analysis of elastic electron scattering up to high $q^2$ using the Fourier Bessel expansion \cite{Off91} and extrapolated with a Whittaker function for very large radii. The $rms$ radius is taken as the quoted best value with 2.478(9)\ fm \cite{Off91}. We assume the same densities for protons and neutrons.

\begin{figure}[t]
\includegraphics[width=0.49\columnwidth]{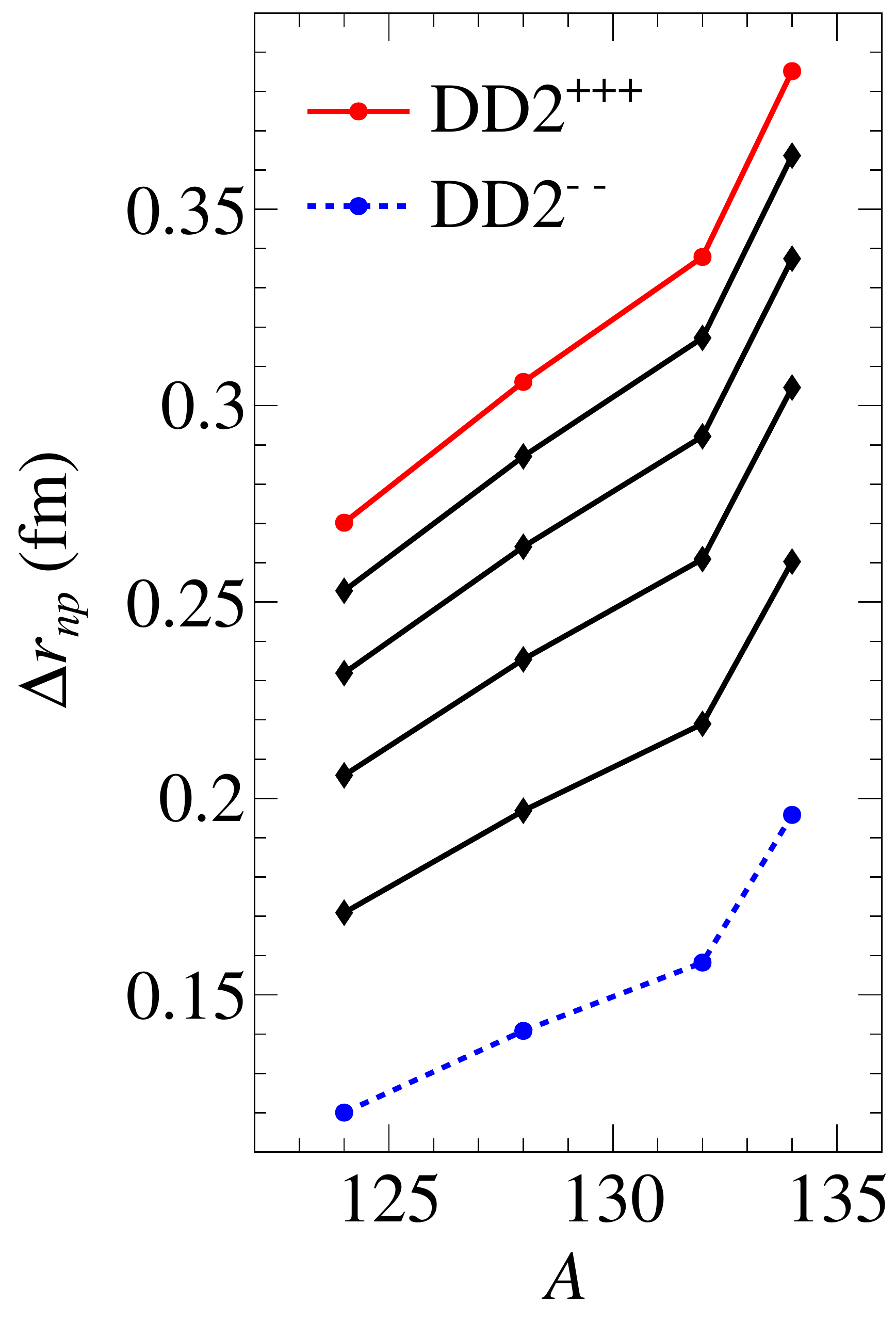}
\includegraphics[width=0.49\columnwidth]{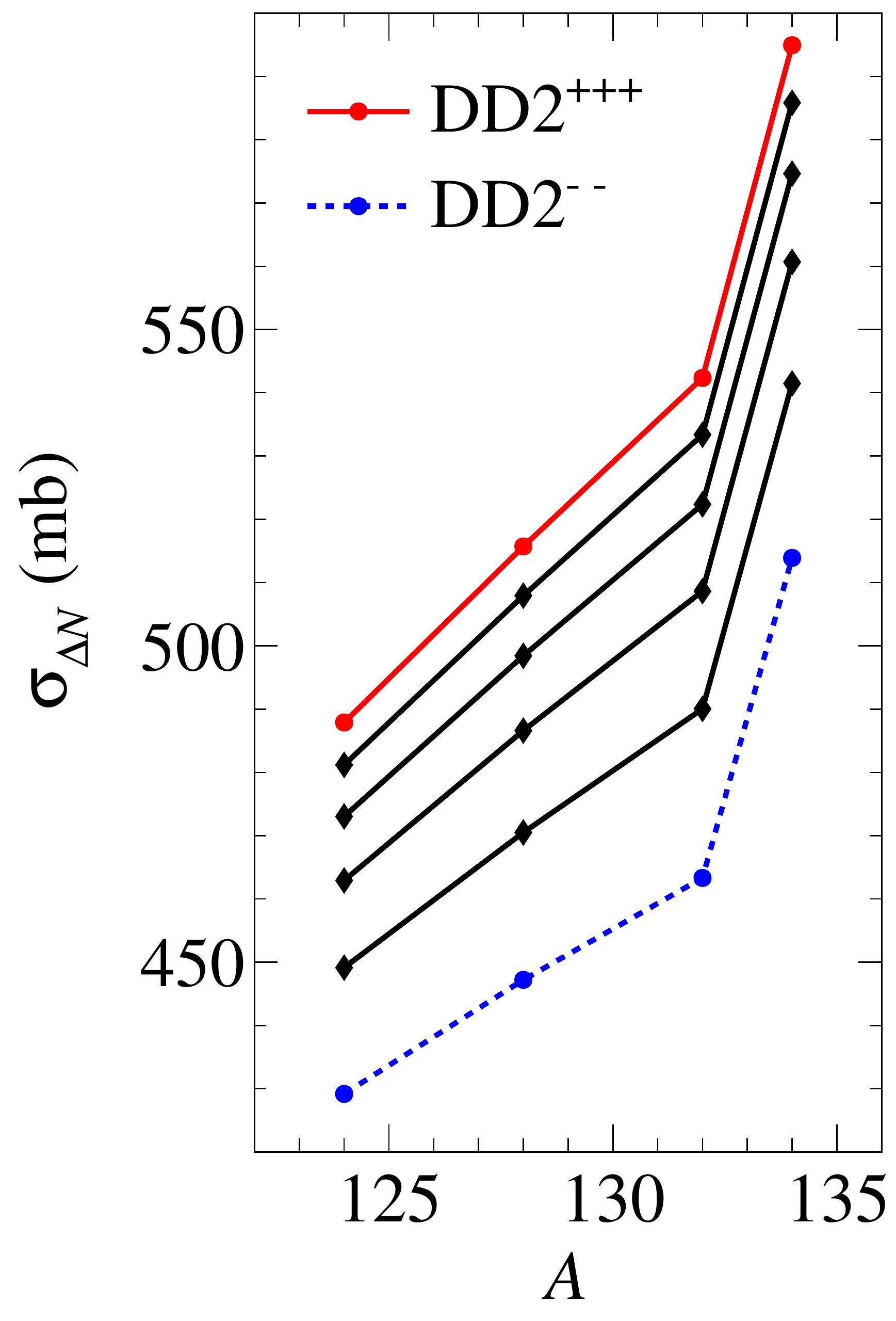}
\caption{(color online) Neutron-skin thickness $\Delta r_{np}$ (left) and corresponding neutron-removal cross sections $\sigma_{\Delta N}$ (right) for Sn isotopes as predicted by RMF calculations based on variations \cite{Typ14} of the DD2 interaction \cite{Typ10}. The slope parameter $L$ has been systematically varied from 25 MeV (DD2$^{--}$) to 100 MeV (DD2$^{+++}$) \cite{Typ14}. }
\label{fig2}
\end{figure}

In order to estimate the sensitivity of $\sigma_{\Delta N}$ with respect to $\Delta r_{np}$ and $L$, we calculate the cross sections using theoretical density distributions from RMF calculations. We have chosen for this sensitivity test the DD2 interaction which has been developed in Ref.\,\cite{Typ10}  and systematically varied in the slope parameter $L$ optimizing the isovector parameters by a fit to nuclear properties including masses and radii \cite{Typ14}. The same protocol as for the DD interaction \cite{Typ05} has been used. The left frame in Fig.\,\ref{fig2} shows the predicted neutron-skin thicknesses for the tin isotopes. The different interactions range from $L$ values of 25\,MeV (DD2$^{--}$) to 100\,MeV (DD2$^{+++}$) and predict accordingly different values of $\Delta r_{np}$ between 0.15 and 0.34\,fm for $^{132}$Sn. This causes a corresponding change in $\sigma_{R}$ from around 2550 to 2610\,mb, i.e., 2.5\%. The quantity, which is most sensitive to $\Delta r_{np}$ is $\sigma_{\Delta N}$ shown in the right panel of Fig.\,\ref{fig2}. A change from 460 to 540\,mb is visible for $^{132}$Sn, i.e., a change of almost 20\%. That is, $\sigma_{\Delta N}$, has a larger potential to tightly constrain $L$ and is less sensitive to imperfections of the reaction theory.

\begin{figure}[t]
\includegraphics[width=0.99\columnwidth]{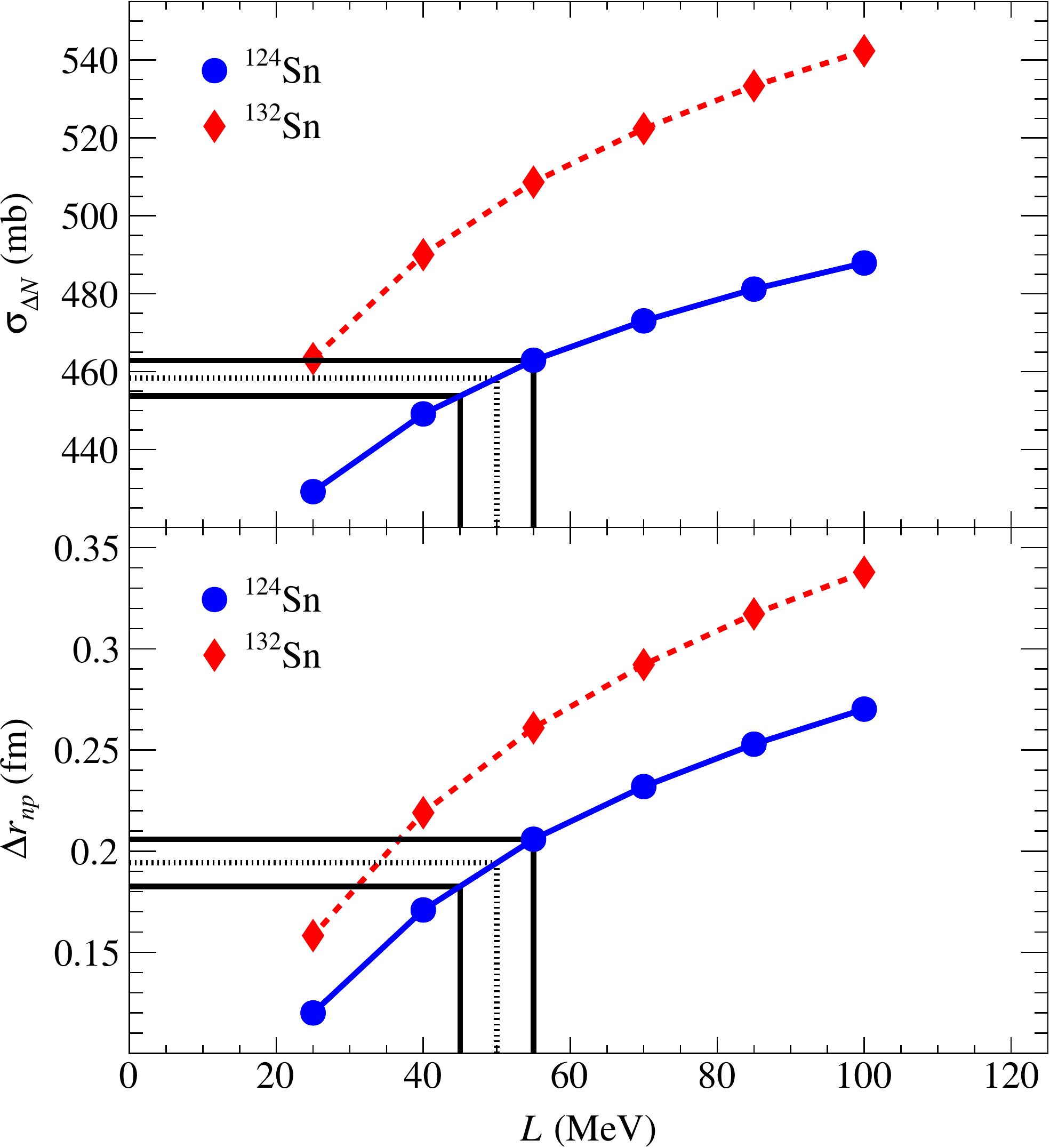}
\caption{(color online) Relation between $\sigma_{\Delta N}$ (top) and $\Delta r_{np}$ (bottom) with the slope parameter $L$ calculated based on RMF theory for $^{124}$Sn and $^{132}$Sn. The lines indicate the sensitivity of the observables to $L$ for a $L$-range of 10 MeV.}
\label{fig3}
\end{figure}

Fig.\,\ref{fig3} displays the correlation between the $L$ value chosen in the DD2 interaction and $\Delta r_{np}$ calculated for $^{124}$Sn and $^{132}$Sn. With this particular interaction, a change of $L$ by $\pm 5$ MeV changes the calculated skin in $^{124}$Sn by around $\pm 0.01$ fm. The same change in $L$ causes a change in $\sigma_{\Delta N}$ by around $\pm 5$ mb, i.e., around $\pm 1$\%. This means that with a determination of $\sigma_{\Delta N}$ with a 1\% accuracy both experimentally and theoretically, the theoretical limit for constraining $L$ via comparison with DFT as discussed earlier could be reached. The scatter of different relativistic and non-relativistic models with given $L$ for the prediction of $\sigma_{\Delta N}$ is expected to be similar as for $\Delta r_{np}$ analyzed in Ref. \cite{Roc11}, i.e., around 10 MeV in $L$. A full analysis with many relativistic and non-relativistic models will follow in a forthcoming article. It should be noted that the dependence of the cross section on $L$ is steeper for the more neutron-rich nucleus $^{132}$Sn providing thus an even higher sensitivity. 

The remaining key point in order to relate DFT and the corresponding symmetry-energy parameters with the measured cross sections is the accuracy of the reaction theory and the assessment of its uncertainty. In order to do so, we start with a parameter-free calculation that enables systematic improvements and tests as well as the quantification of its uncertainty. We compare our calculations to data available in the literature and propose sensitive measurements that will uncover any discrepancy between experiment and theory.

Nuclear fragmentation in high-energy collisions  is usually studied via two completely disconnected theoretical models: (a) primary fragment production due to multi-nucleon removal via nucleon-nucleon collisions (as described above), followed by (b) secondary fragments produced via nucleon evaporation due to the energy deposit in primary fragments. The second step is highly model dependent, usually based on the Hauser-Feshbach theory of  compound-nucleus decay. The method used in this work does not require a consideration of the nuclear evaporation step, as the total neutron and charge removal cross sections basically account for the completeness of the sum over all decay channels. It is important to note that proton or charged-particle evaporation is negligible in the cases of $^{124}$Sn and heavier tin isotopes as discussed here. For example, the calculated $\sigma_{\Delta N}$ for the production of primary fragments for  580\,MeV/nucleon $^{124}$Sn incident on $^{12}$C is $\sigma_{\Delta N}=485.6$~mb, while the same cross section calculated after the evaporation stage, using traditional parameters in the Hauser-Feshbach formalism, is  $\sigma_{\Delta N}=483.4$~mb, i.e., less than 0.5\% of the neutron-removal cross section is transferred to the charge-changing cross section after the primary reaction stage. For more neutron-rich tin isotopes, the effect becomes even smaller. Changes in the input parameters used in the Hauser-Feshbach calculations are by no means able to increase this effect appreciably. 

In addition to fragmentation processes induced by nucleon-nucleon collisions as discussed so far, the projectile can loose a nucleon after inelastic excitation of collective states in the continuum such as giant resonances. For heavy neutron-rich nuclei as considered here, this process contributes almost exclusively to the neutron-removal channel and to the total interaction cross section, which we define as the sum of the two processes, $\sigma_{I}=\sigma_{R}+\sigma_{inel}$. We will not attempt to calculate  $\sigma_{inel}$ which contains a nuclear and an electromagnetic contribution and its interference. We estimate the contribution of $\sigma_{inel}$ to be of the order of 1\% in case of $^{12}$C + $^{12}$C at energies above 600 MeV/nucleon, while it can reach values around 100\,mb in case of $^{132}$Sn+$^{12}$C \cite{Schr15,Schi17} corresponding to 4\% or 20\% of $\sigma_{I}$ or $\sigma_{\Delta N}$, respectively (the probability for charged particle evaporation is extremely low for neutron-rich heavy nuclei, see above). Since it is the neutron-removal cross section providing the sensitivity to the neutron skin, this contribution has to be known with an uncertainty $<$5\% in order to reach the required accuracy. While this seems impossible to reach presently with reaction theory, it is possible with state-of-the-art kinematical complete experimental measurements to separate this contribution and determine its cross section. The fact that the angular distributions of neutrons are very different for the two processes can be used to separate the contributions experimentally. Since evaporated neutrons (with typical energies around 2\,MeV in the rest frame of the projectile) are kinematically boosted to forward direction at high beam energy, they can be detected around $0\degree$  with beam velocity. The angular distribution covers typically a range of $0$ to $5\degree$, while neutrons stemming from a nucleon-nucleon collision have a broad angular distribution ranging from 0 to $90\degree$  with a maximum at $45\degree$. The overlap region is thus negligible. 

Since the calculation of the primary process of nucleon-nucleon collisions remains the only significant step towards relating $\sigma_{\Delta N}$  with $\Delta r_{np}$ or $L$, the reaction model and its uncertainty reduces to the eikonal theory described above. In order to test the performance of our model we start with the case of the symmetric system $^{12}$C + $^{12}$C, where experimental information on $\sigma_R$ is available, using the eikonal approximation in its simplest form as given in Eqs.\,(1) and (2). The known free nucleon-nucleon cross sections and the densities serve as only input to the reaction theory. We omit any adjusted additional energy-dependent parameters as often done \cite{Ray79,Hor07,Kan16}, which would mask deficiencies from the optical limit eikonal approximation and thus would preclude a systematic improvement of the theory and a quantitative assessment of its uncertainty. The results for the total reaction cross section as a function of the laboratory beam energy are shown as black triangles in the lower panel of Fig.\,\ref{fig1}. We notice that the calculated cross sections overestimate the experimental data for energies larger than 200 MeV/nucleon. We expect that in-medium effects are the dominant reason for deviations at high beam energies. A large fraction of this deviation can be indeed accounted for when Pauli blocking is taken into account (red diamonds). Pauli blocking was calculated as in Ref.\,\cite{BC10}. Still, the high-energy data point at around 950\,MeV/nucleon is overestimated, although by only about 2\%. Below 400 MeV/nucleon, where the data start to deviate strongly from the calculation, we expect that effects beyond the eikonal approach start to play an increasingly large role. According to the work of Ref.\,\cite{Tak09}, the effect of Fermi motion becomes important in this energy regime and yields an increase of the cross sections. We will thus not consider energies below 400\,MeV/nucleon. It should be noted that in the most relevant energy region ($400-1200$ MeV/nucleon) only three data points exist, and none in the important region between 400 and 800 MeV/nucleon, where the cross section increases as a function of energy. Since deviations from the eikonal approximation like in-medium and higher-order effects should depend on the beam energy, high-precision data covering this energy range with $<1\%$ accuracy are thus of utmost importance for both, a stringent test and further development of the reaction theory, as well as for the quantification of its uncertainty. 

\begin{figure}[t]
\includegraphics[width=0.99\columnwidth]{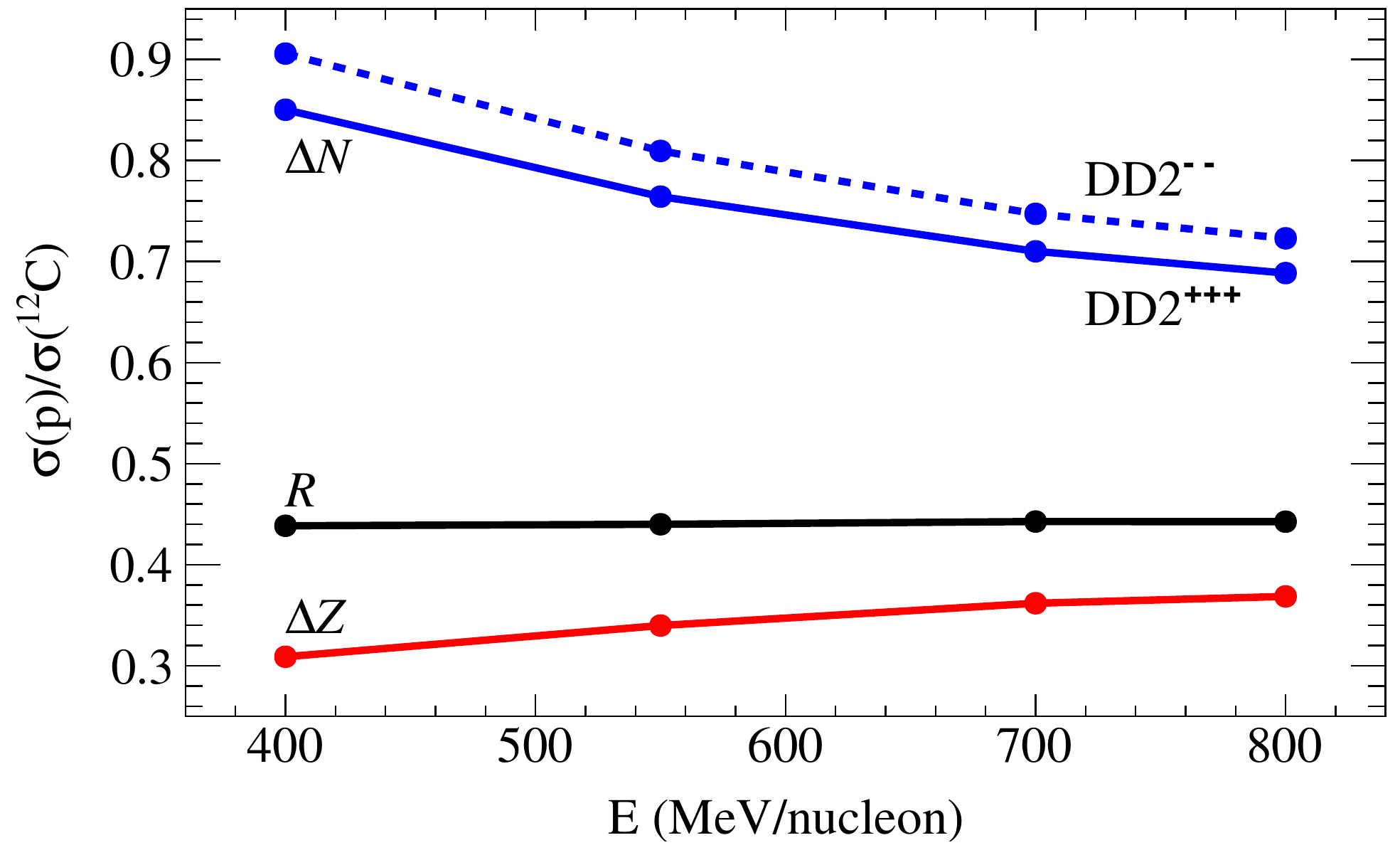}
\caption{(color online) Ratios of $\sigma_R$, $\sigma_{\Delta Z}$ and $\sigma_{\Delta N}$ for $^{134}$Sn projectiles incident on proton and $^{12}$C targets as a function of the bombarding energy.}
\label{fig4}
\end{figure}

Further sensitivity can be achieved by varying the reaction target. Since the $np$ and $pp$ cross sections have a very different energy dependence, see Fig.\,\ref{fig1}, we expect a corresponding change in the ratio of neutron-removal to charge-changing cross sections as a function of energy. This effect should be most pronounced for a proton target since the proton target probes the neutron skin exclusively via $pn$ reactions, while charge changing is exclusively related to $pp$ reactions. There are additional subtle effects as the proton has a non-negligible chance to swift through the nucleus without knocking out a nucleon in contrast to $^{12}$C targets that all but probes the surface of the nucleus. The consequences become evident in Fig.\,\ref{fig4} where we plot the ratios of $\sigma_R$, $\sigma_{\Delta Z}$ and $\sigma_{\Delta N}$ for $^{134}$Sn projectiles incident on proton targets with those on $^{12}$C targets as a function of the bombarding energy. Whereas no energy dependence is seen for  the $\sigma_R({\rm p})/\sigma_R(^{12}{\rm C})$ target ratio, the charge-changing $\sigma_{\Delta Z}({\rm p})/\sigma_{\Delta Z}(^{12}{\rm C})$ and $\sigma_{\Delta N}({\rm proton})/\sigma_{\Delta N}(^{12}{\rm C})$ target ratios clearly display a laboratory energy dependence. The energy dependence of the ratio and the fact that the ratio is significantly larger for $\sigma_{\Delta N}$ is related to the strong energy dependence of the $pp$ cross section (see Fig.\,\ref{fig1}) yielding a substantial proton survival probability with the proton target at 400\,MeV/nucleon and thus a larger $\sigma_{\Delta N}$, while this effect becomes much smaller for energies of 800 MeV/nucleon and above. The energy dependence of the ratio for $\sigma_{\Delta N}$ provides thus a very sensitive test to the reaction theory if measured accurately. Moreover, both ratios  for $\sigma_R$ and $\sigma_{\Delta Z}$ have negligible dependence on the neutron skin while the ratio for $\sigma_{\Delta N}$ shows an explicit dependence on $\Delta r_{np}$ as evidenced by the use of the DD2$^{+++}$ and DD2$^{--}$ RMF interactions. Since the $rms$ radius of the charge distribution is known, the charge-changing cross sections for proton and carbon targets as a function of bombarding energy can serve as an additional crucial test on the accuracy of the predicted cross sections. 

In summary, in this Letter we have proposed a new and robust technique to study the evolution of the neutron-skin thickness in nuclei far from stability. The idea is to use hadronic interactions in relativistic heavy-ion collisions and measurements of total neutron-removal cross sections. We have shown that several experimental variations like using different targets, specific ranges of bombarding energies, or a large variety of radioactive nuclear beams can be used to track the sensitivity of the measurements with the neutron-skin thickness, to prove the validity of the parameter-free reaction model, and its uncertainty and to guide a systematic improvement of it. With this, the method devised here is the most promising to determine the neutron skin of unstable very neutron-rich heavy nuclei and to constrain the density dependence of the EoS of neutron-rich matter with quantified uncertainties and model dependencies. Since the proposed measurements are already possible to be performed in existing radioactive-beam facilities and with newly constructed detectors, we hope that we can get much closer to understand the role of the symmetry energy in nuclei and in neutron stars in the near future.  

This project was supported by the BMBF via project 05P15RDFN1, through the GSI-TU Darmstadt cooperation agreement, the DFG via the SFB1245, and in part by the U.S.\ DOE grant DE- FG02-08ER41533 and the U.S.\ NSF Grant No.\ 1415656. We thank HIC for FAIR for supporting visits (C.A.B.) to the TU Darmstadt.

\end{document}